\documentclass[sigconf]{acmart}
\setcopyright{none}
\usepackage{amsmath}
\usepackage{booktabs}
\usepackage{graphicx}

\acmConference{}{}{}
\acmBooktitle{}
\acmYear{}
\acmISBN{}
\acmDOI{}
\begin{document}

\title{Modeling User Exploration Saturation: When Recommender Systems Should Stop Pushing Novelty}
\author{Enock O. Ayiku}
\affiliation{%
  \institution{University of Massachusetts Boston}
  \city{Boston}
  \state{MA}
  \country{USA}
}
\email{enock.ayiku001@umb.edu}

\author{Onyeka Emebo}
\affiliation{%
  \institution{Virginia Tech}
  \city{Blacksburg}
  \state{VA}
  \country{USA}
}
\email{onyeka@vt.edu}

\author{Evelyn Osei}
\affiliation{%
  \institution{Kansas State University}
  \city{Manhattan}
  \state{KS}
  \country{USA}
}
\email{evelynnanaosei@gmail.com}
\begin{abstract}
	Fairness-aware recommender systems often mitigate bias by increasing exposure to
under-represented or long-tail content, commonly through mechanisms that promote
novelty and diversity. In practice, the strength of such interventions is typically controlled using global hyperparameters, fixed regularization weights, heuristic caps, or offline tuning strategies. These approaches implicitly assume that a single level of exploration is appropriate across users, contexts, and stages of interaction. In this work, we study exploration saturation as a user-dependent phenomenon arising from fairness- and novelty-driven recommendation strategies. We define exploration saturation as the point at which further increases in exploration no longer improve user utility and may instead reduce engagement or perceived relevance. Rather than proposing a new fairness-aware algorithm or optimizing a specific objective, we empirically analyze how increasing exploration affects users across varied recommendation models. Through longitudinal experiments using MovieLens-1M and Last.fm datasets, our results indicate that fairness-induced exploration exhibits diminishing or non-monotonic returns and varies substantially across users. In particular, users with limited interaction histories tend to reach saturation earlier, suggesting that uniform fairness or novelty pressure can disproportionately disadvantage certain users. These findings reveal a trade-off between fairness and user experience, suggesting that recommendation systems should adapt not only to relevance but also to the amount of fairness-driven exploration applied to individual users.
\end{abstract}


\maketitle

\section{Introduction}
Recommender systems play a central role in shaping what users consume, discover,
and engage with \cite{hazrati2024choice, knijnenburg2012explaining}. Beyond accuracy, modern recommendation research \cite{zhao2024fairness,abdollahpouri2017controlling,ge2021towards,kaminskas2016diversity,wu2022multi} has
increasingly emphasized fairness, diversity, equitable exposure, serendipity, novelty, and coverage motivated
by concerns over popularity bias \cite{zhu2021popularity}, filter bubbles \cite{Areeb2023Filter}, and the systematic
under-representation of long-tail items \cite{Yin2012Challenging}.
A common strategy across fairness-aware and bias-mitigation approaches is to increase exposure to under-represented items by promoting novelty or diversity in recommendation
lists.
While such interventions are well-intentioned, they implicitly increase the
degree of exploration imposed on users. In practice, fairness-driven
recommendation often requires users to engage with unfamiliar content, new
genres, or less popular items. Existing work\cite{wu2022multi,karimi2023provider} typically assumes that increasing
exploration to enhance fairness is uniformly beneficial, or at least
harmless, and focuses on balancing fairness objectives against traditional
accuracy metrics. However, this assumption overlooks an important behavioral
question: how much exploration can users tolerate before these interventions
begin to degrade user experience?

Real users interact with recommender systems under cognitive and behavioral
constraints. Excessive novelty can increase cognitive load, reduce perceived
relevance, and lead to disengagement or reversion to familiar content\cite{sweller1988cognitiveload}. These
effects suggest that exploration may not be an unbounded good, but rather a
bounded process with user-specific limits. Despite this, prior work rarely
models when exploration ceases to be beneficial, nor how user tolerance for
fairness-induced novelty varies across individuals.

In this work, we study \emph{exploration saturation} as a fundamental but
underexplored phenomenon in fairness-aware recommendation. We define exploration
saturation as the point at which increasing exploration no longer improves user
utility and may instead lead to stable or declining engagement or relevance.
Importantly, we do not propose a new fairness-aware recommendation algorithm.
Rather, our goal is to empirically characterize how users respond to increasing
exploration across different recommendation models and to identify whether
uniform exploration pressure can lead to unequal user impact.
Through longitudinal experiments\cite{Hu2022A} on MovieLens-1M\cite{harper2015movielens} and Last.fm\cite{celma2010lastfmzenodo} dataset, we analyze how
user utility responds to increasing levels of exploration induced by different
recommendation strategies. Our analysis reveals that exploration often exhibits
non-monotonic or diminishing returns, with saturation occurring at different
levels across users, models, and domains. In particular, users with limited interaction histories tend to reach saturation earlier, suggesting that uniform fairness or novelty promotion can disproportionately disadvantage certain user groups.

This perspective reframes fairness not as a quantity to be universally maximized, but as a user-dependent intervention that must be carefully balanced against individual tolerance for exploration and experimentation. By highlighting the limits of fairness-driven novelty, our work motivates adaptive approaches that personalize not only relevance but also the degree of exploration introduced by fairness interventions.
Unlike classical accuracy–diversity or exploration–exploitation trade-offs, exploration saturation does not describe a global optimization frontier or a model miscalibration effect. Instead, it characterizes a user-experienced limit at which additional exploration ceases to provide marginal benefit, even when aggregate relevance metrics remain stable. In our experiments, saturation frequently manifests as utility plateaus or instability rather than sharp performance collapse, indicating that it cannot be reduced to conventional accuracy degradation or poor hyperparameter tuning. This distinction allows exploration saturation to surface behavioral constraints that remain invisible under standard trade-off analyses
\paragraph{Contributions.}
This work makes the following contributions:
\begin{itemize}
    \item We introduce \emph{exploration saturation}\cite{radlinski2008too, valko2013finite, li2010contextual} as a user-dependent,
    behavioral phenomenon that characterizes the limits of fairness- and
    novelty-driven recommendation, reframing exploration as a bounded,
    user-experienced process rather than a uniformly beneficial objective.

    \item We propose a diagnostic framework for analyzing
    exploration--utility relationships that identifies saturation regimes
    through aggregate trends, marginal utility effects, and user-level
    heterogeneity, without requiring explicit exploration tuning or new
    optimization objectives.

    \item Through longitudinal experiments on MovieLens-1M and Last.fm, we
    empirically demonstrate that increasing exploration often yields
    diminishing, unstable, or negative utility returns, with saturation
    occurring at different levels across recommendation models.

    \item We show that uniform exploration strategies impose unequal user
    impact: users with limited interaction histories consistently reach
    exploration saturation earlier, highlighting an underexplored
    fairness--utility trade-off that is invisible to population-level metrics.
\end{itemize}

\section{Related Work}

\subsection{Exploration and Novelty in Recommendation}

Exploration-exploitation trade-offs are a foundational topic in recommender
systems and online decision-making. Bandit-based and reinforcement learning\cite{li2010contextual,zhou2010solving,swaminathan2017counterfactual}.
approaches introduce exploration to improve long-term learning \cite{Yoo2025Continual, Urdaneta-Ponte2021Lifelong}, reduce
uncertainty, and avoid premature convergence, typically assuming that exploration
is beneficial until sufficient information is acquired. Similarly,
novelty and diversity-aware recommendation methods explicitly promote exposure
to unfamiliar or less popular items to counteract popularity bias\cite{liu2023mitigating} and filter
bubbles.

Across these approaches, exploration is commonly treated as an optimization
variable controlled through global hyperparameters, re-ranking strategies, or
reward trade-offs. Evaluation therefore, focuses on identifying optimal operating
points that balance short-term accuracy with long-term gains. While effective
for learning and coverage, these methods implicitly assume that increased
exploration remains desirable as long as performance does not sharply degrade.

In contrast, our work does not seek to optimize exploration. Instead, we
empirically examine the limits of exploration as experienced by users and study
when additional exploration ceases to provide utility, even before traditional
performance collapse occurs.

\subsection{User Modeling and Behavioral Constraints}

User modeling research\cite{adomavicius2011context,ricci2015recommender} has long emphasized that users differ in preferences,
expertise, attention, and cognitive capacity. Prior studies have examined
preference drift \cite{Sritrakool2021Personalized}, session-level intent \cite{zhang2025adaptive}, choice overload \cite{peng2021does}, and cognitive burden,
demonstrating that excessive novelty or choice can reduce engagement and
satisfaction.

Despite these insights, recommender system evaluations rarely incorporate
behavioral constraints when assessing exploration strategies. Users are often
implicitly assumed to tolerate increasing novelty or diversity as long as
aggregate relevance metrics remain stable, particularly in offline evaluations.
As a result, the behavioral cost of exploration is largely unmeasured.

Our work bridges this gap by operationalizing exploration tolerance as a
measurable, user-dependent phenomenon. By analyzing how user utility evolves as
exploration increases, we reveal substantial heterogeneity in user response that
is not captured by aggregate metrics.

\subsection{Fairness and Unequal User Impact}

Fairness-aware recommender systems aim to mitigate biases such as popularity skew
and under-exposure of minority items by adjusting exposure distributions. Many
such approaches promote long-tail \cite{chen2020esam} or under-represented content through increased
novelty or diversity, implemented via regularization\cite{kiswanto2018fairness}, re-ranking \cite{10.1145/3298689.3347000}, or constraints.
While these methods improve fairness at the catalog or population level,
evaluation primarily focuses on trade-offs between accuracy and exposure-based
fairness metrics. Less attention is paid to how fairness-driven exploration is
experienced by individual users.
Recent work on bias mitigation has begun to highlight unequal user impact in recommender systems,
showing that interventions can benefit some users while harming others. However,
exploration itself is rarely isolated as a source of heterogeneous user burden.
Our work complements this literature by empirically demonstrating that uniform
exploration pressure can impose unequal experiential costs across users.
\section{Exploration Saturation}

\subsection{Motivation}

Fairness- and novelty-driven recommendation strategies frequently increase user
exposure to unfamiliar content. While such exploration can improve discovery and
mitigate bias, it also imposes cognitive and relevance-related costs on users.
These costs suggest that exploration may not be an unbounded good, but rather a
bounded process with user-specific limits.
Existing recommender system evaluations rarely model user-specific limits on exploration explicitly. Instead, exploration is typically increased until aggregate performance metrics decline, implicitly assuming that exploration saturation only occurs when the system as a whole begins to fail. This evaluation practice overlooks earlier, user-level signals of diminishing or unstable benefit, where additional exploration no longer improves outcomes and may already be harmful for certain users, even though overall metrics remain stable.

\subsection{Definition of Exploration Saturation}

Let $u \in \mathcal{U}$ denote a user and $t$ a recommendation step. Let
$E_{u,t}$ represent the level of exploration induced at step $t$, measured using
a novelty- or diversity-based metric (e.g., recommendation entropy or fraction
of unseen items). Let $U_{u,t}$ denote a proxy for user utility, for example, how likely the user is to interact with the next recommended items or continue their session.

We define \emph{exploration saturation} for user $u$ as the regime in which
increases in exploration no longer yield positive marginal utility. Formally,
user $u$ is said to be saturated at exploration level $E$ if:
\[
\frac{\partial \mathbb{E}[U_{u} \mid E_{u}=E]}{\partial E} \leq 0.
\]

This definition captures multiple saturation behaviors:
(i) declining utility, where additional exploration harms engagement;
(ii) plateauing utility, where exploration yields no further benefit; and
(iii) unstable utility, where gains become inconsistent or highly variable.

Importantly, saturation does not require absolute performance degradation.
Rather, it reflects the exhaustion of marginal benefit from exploration for a
given user.

\subsection{Detecting Saturation Points}

In practice, direct estimation of derivatives is infeasible in offline
evaluation. We therefore detect saturation through discrete approximations over
exploration regimes.

We partition recommendation events into ordered exploration quantiles
$\{Q_1, \dots, Q_K\}$ based on $E_{u,t}$. Let $\bar{U}_{u}(Q_k)$ denote the average
utility for user $u$ within quantile $Q_k$. The marginal effect of exploration is
approximated as:
\[
\Delta U_{u}(Q_k) = \bar{U}_{u}(Q_k) - \bar{U}_{u}(Q_{k-1}).
\]

We identify saturation when $\Delta U_{u}(Q_k) \leq 0$ for consecutive quantiles,
or when $\Delta U_{u}(Q_k)$ converges toward zero with increasing variance. This
approach allows us to capture both sharp and gradual saturation effects without
assuming monotonic utility trends.

At the population level, we analyze the distribution of user-specific saturation
points to characterize heterogeneity in exploration tolerance. This enables
comparison across user groups (e.g., short- vs.\ long-history users), models, and
datasets, and reveals unequal user impact arising from uniform exploration
strategies.

\section{Experimental Setup}

This section describes the datasets, recommendation models, exploration metrics,
user utility proxies, and experimental protocol used to empirically study
exploration saturation in fairness and novelty-driven recommendation. Our experimental design is intentionally diagnostic rather than optimization-driven, with our goal to
characterize how increasing exploration affects users, rather than
to identify a single optimal configuration.

\subsection{Datasets}

We conduct experiments on two publicly available datasets that differ
substantially in domain, sparsity, and interaction patterns. These differences
allow us to examine whether exploration saturation is a general behavioral
phenomenon or a dataset-specific artifact.

\textbf{MovieLens-1M}\cite{harper2015movielens} is a movie recommendation dataset containing approximately
one million ratings from 6,040 users on 3,706 movies. User interaction histories
in MovieLens-1M are relatively dense, and user preferences tend to be stable
over time. This dataset provides a controlled setting in which users are often
familiar with a large portion of the catalog, making exploration comparatively
low-risk.

\textbf{Last.fm}\cite{schedl2016lfm1b} captures music listening behavior with long but sparse user
interaction sequences and a substantially larger and more dynamic item space.
Compared to MovieLens, Last.fm exhibits higher item turnover, stronger popularity
skew, and greater preference heterogeneity. These characteristics make novelty
and exploration more pronounced, and potentially more taxing for users.

Table~\ref{tab:dataset_stats} summarizes the key statistics of both datasets.

\begin{table}[t]
\centering
\caption{Summary statistics of the datasets used in the experiments.}
\label{tab:dataset_stats}
\begin{tabular}{lcccc}
\toprule
Dataset & Users & Items & Interactions & Domain \\
\midrule
MovieLens-1M & 6,040 & 3,706 & 1,000,209 & Movies \\
Last.fm & 1,000 & 17,632 & 19,150,868 & Music \\
\bottomrule
\end{tabular}
\end{table}

For both datasets, interactions are temporally ordered for each user. We follow
a standard sequential evaluation protocol, using earlier interactions for
training and later interactions for evaluation \cite{boka2024survey,wang2022sequential}. This setup allows us to analyze
how user utility evolves as users are repeatedly exposed to recommendations with
varying exploration intensity.

\subsection{Recommendation Models}

To isolate exploration saturation as a behavioral phenomenon rather than also an
algorithm-specific artifact, we evaluate representative recommendation models
that implicitly induce different levels of novelty and exploration:

\textbf{MostPopular}\cite{celma2008music} recommends globally popular items and serves as a
low-exploration baseline. Because it relies solely on aggregate popularity, it
provides stable but minimally novel recommendations.

\textbf{BPR-MF}\cite{rendle2009bpr} employs Bayesian Personalized Ranking with matrix factorization,
introducing personalization while remaining relatively conservative in its
exposure to unfamiliar items.

\textbf{Neural Collaborative Filtering (NCF)}\cite{he2017ncf} models non-linear user--item
interactions through neural architectures. Its increased representational
capacity often leads to greater exposure of less popular or less familiar items,
implicitly increasing exploration pressure.

\textbf{LightGCN}\cite{he2020lightgcn} propagates user and item representations through the
interaction graph. Neighborhood aggregation can amplify exposure beyond direct
historical interactions, frequently resulting in higher diversity and novelty.

These models span popularity-based, latent-factor, neural, and graph-based
recommendation paradigms \cite{celma2008music,rendle2009bpr,he2017ncf,he2020lightgcn}.
Importantly, we do not explicitly tune novelty or fairness parameters. Instead,
variation in exploration arises naturally from differences in model structure and
inductive bias.
We intentionally focus on exploration that arises implicitly from model inductive biases rather than from explicit fairness or diversity controls. This design isolates exploration as a behavioral phenomenon that emerges even in standard recommendation pipelines, independent of additional optimization objectives. Importantly, if exploration saturation appears under these baseline conditions, it represents a lower bound on user tolerance and is likely to persist—or intensify—under stronger, explicitly enforced fairness or novelty interventions.

\subsection{Exploration Metrics}
We quantify exploration using multiple complementary metrics computed over
top-$k$ recommendation lists. These metrics capture different dimensions of
novelty and diversity that are commonly leveraged by fairness-aware
recommendation strategies:

\paragraph{Exploration Metrics.}
We quantify exploration using multiple complementary metrics computed over top-$k$ recommendation lists. These metrics capture distinct dimensions of novelty and diversity that are commonly leveraged by fairness-aware recommendation strategies.

\textbf{Recommendation Entropy} captures diversity by computing the entropy of recommended items across semantic clusters, reflecting how evenly recommendations are distributed across content categories.

\subsection{User Utility Proxies}

Because explicit user satisfaction signals are unavailable, we approximate user
utility using observable behavioral proxies that reflect engagement and
relevance:

\textbf{Next-interaction Hit Rate} measures the proportion of users for whom the item they actually interacted with next appears in the model’s top-K recommendations, given their prior interaction history.

\textbf{Session continuation} captures whether users continue interacting after
receiving recommendations, reflecting sustained engagement.

These proxies do not assume monotonic improvement with increased exploration and
are explicitly used to detect diminishing or negative utility returns.
While these proxies do not directly measure subjective experience, they provide observable behavioral signals that are widely used to infer engagement and relevance, allowing us to identify when additional exploration ceases to yield reliable behavioral benefit rather than to make claims about internal cognitive states.

\subsection{Experimental Protocol}

For each dataset and recommendation model, we generate sequential recommendation
lists for users over time. At each recommendation step, we compute exploration
metrics and corresponding utility proxies. Exploration variation arises
naturally from differences in model behavior rather than from explicit tuning of
exploration or fairness parameters.
We analyze the relationship between exploration and utility at both aggregate
and individual user levels. Aggregate analysis reveals overall trends in how
exploration affects utility, while user-level analysis captures heterogeneity in
exploration tolerance. Exploration saturation is identified when increases in
exploration are associated with diminishing or non-positive marginal changes in
utility.

All analyses are conducted consistently across datasets and models to ensure
comparability.

\section{Results}

This section presents empirical results analyzing how user utility responds to
increasing exploration across datasets and recommendation models. We report
aggregate trends, model-specific behavior, marginal effects, and user-level
heterogeneity using all figures generated from our experiments. Rather than
expecting exploration to uniformly produce performance degradation, our goal is
to characterize when exploration yields gains, when it saturates, and when it
becomes harmful.

\subsection{Aggregate Utility vs. Exploration}
\begin{figure}[t]
\centering
\includegraphics[width=\linewidth]{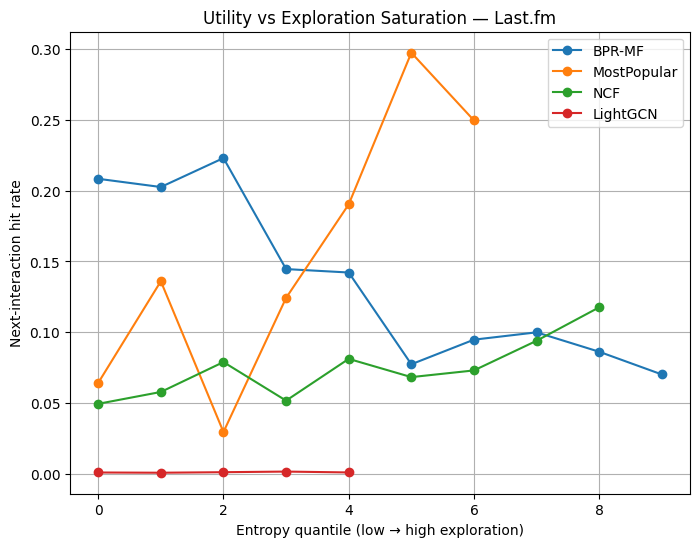}
\caption{Utility vs. exploration on Last.fm across recommendation models.}
\label{fig:lastfm_utility}
\end{figure}

\begin{figure}[t]
\centering
\includegraphics[width=\linewidth]{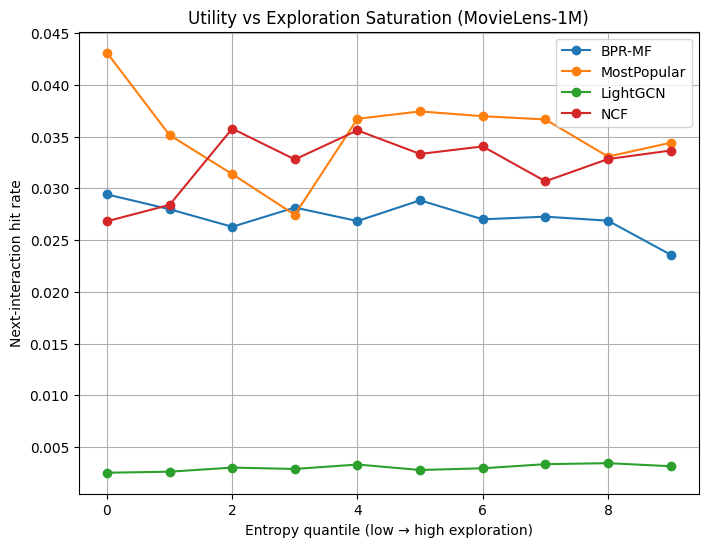}
\caption{Utility vs. exploration on MovieLens-1M across recommendation models.}
\label{fig:movielens_utility}
\end{figure}

We begin by examining aggregate relationships between exploration and user
utility across datasets. Consistent with our definition of exploration
saturation, aggregate saturation is indicated by regions in which increasing
exploration yields non-positive marginal changes in utility, manifesting as
declines, plateaus, or unstable fluctuations rather than monotonic improvement.

Figure~\ref{fig:lastfm_utility} shows utility as a function of
entropy-based exploration quantiles on Last.fm. Several models exhibit clear
non-monotonic behavior. BPR-MF shows a steady decline in utility (next-interaction hit rate) as exploration increases, indicating early exploration saturation. MostPopular exhibits large fluctuations, suggesting unstable utility once exploration exceeds low levels.
NCF improves initially but plateaus and fluctuates at higher exploration
quantiles, indicating exhaustion of marginal benefit rather than continued
gain. LightGCN remains near zero throughout, suggesting that increased
exploration fails to translate into meaningful engagement gains in this domain,
corresponding to saturation through insensitivity rather than collapse.

Figure~\ref{fig:movielens_utility}
presents the same analysis for MovieLens-1M.
Compared to Last.fm, MovieLens exhibits flatter utility curves across models.
While sharp declines are less frequent, utility improvements saturate quickly
and do not increase monotonically with exploration. In this denser domain,
exploration saturation manifests primarily as stagnation and diminishing
marginal gains rather than abrupt performance degradation.

\subsection{Model-Specific Exploration Behavior}

\begin{figure}[t]
\centering
\includegraphics[width=\linewidth]{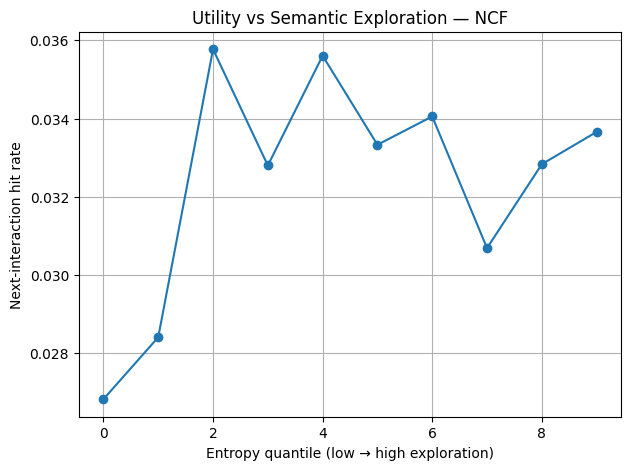}
\caption{Utility vs. semantic exploration for NCF.}
\label{fig:ncf_semantic}
\end{figure}

\begin{figure}[t]
\centering
\includegraphics[width=\linewidth]{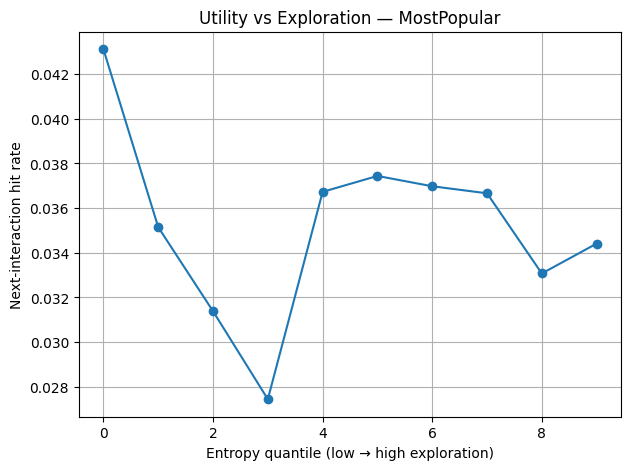}
\caption{Utility vs. exploration for MostPopular on MovieLens-1M.}
\label{fig:mostpop_movielens}
\end{figure}

\begin{figure}[t]
\centering
\includegraphics[width=\linewidth]{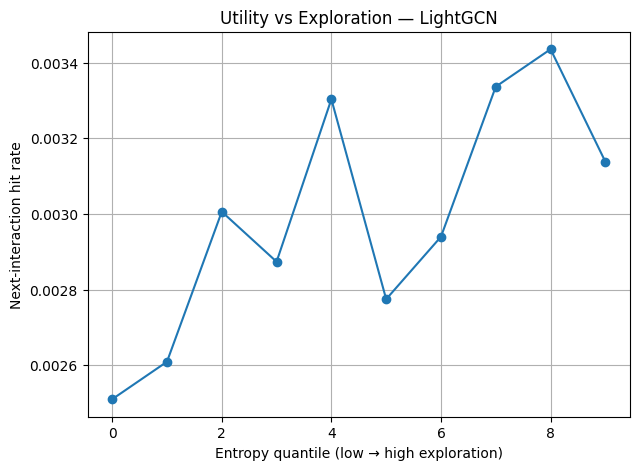}
\caption{Utility vs. exploration for LightGCN.}
\label{fig:lightgcn}
\end{figure}
\begin{figure}[t]
\centering
\includegraphics[width=\linewidth]{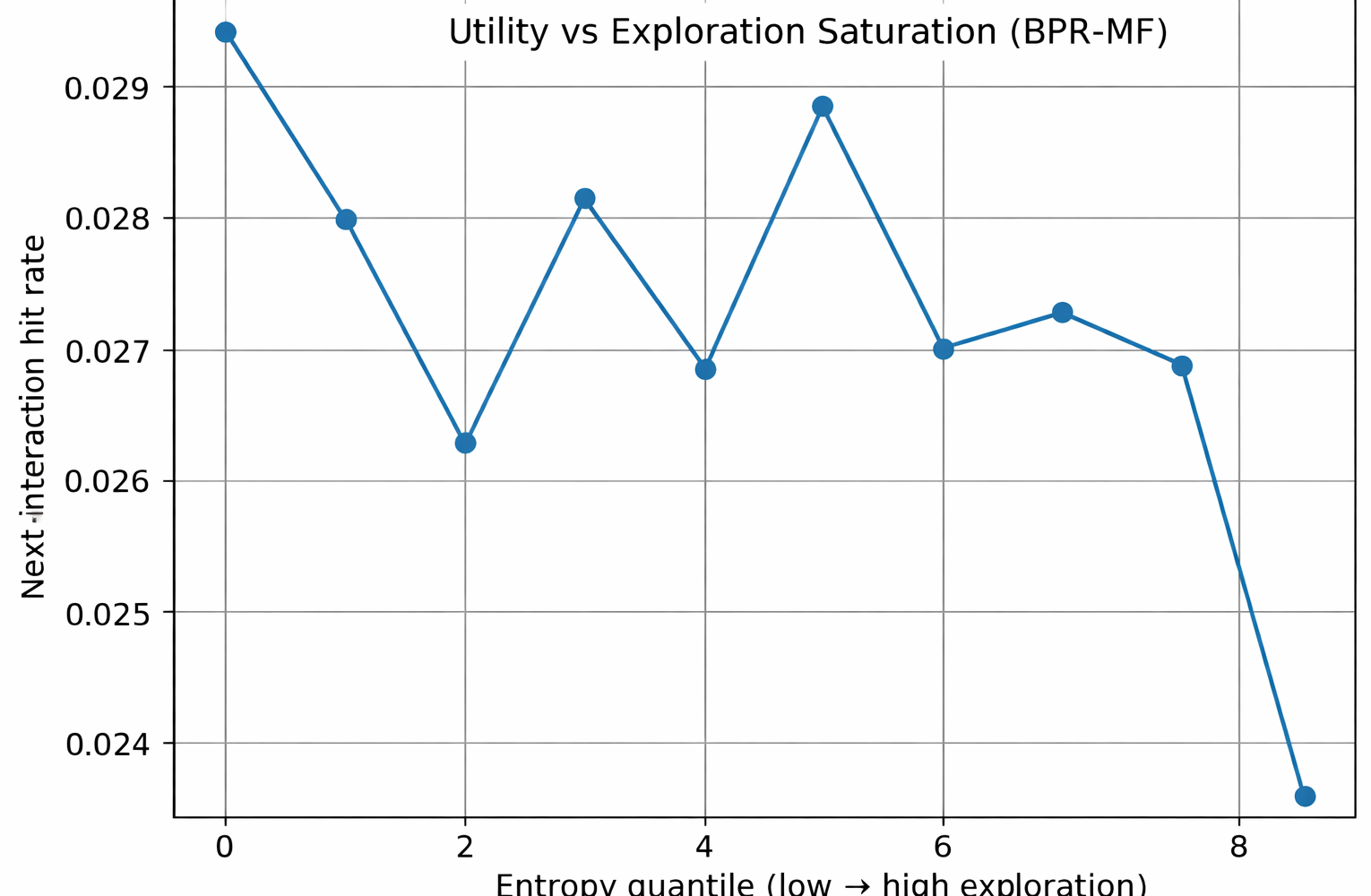}
\caption{Semantic exploration effects for BPR-MF on MovieLens-1M.}
\label{fig:bpr_movielens}
\end{figure}
We next examine model-specific patterns to better understand how different
architectures respond to exploration pressure.

Figure~\ref{fig:bpr_movielens} 
shows semantic exploration effects for BPR-MF on MovieLens-1M. Utility peaks at
low exploration levels and declines at higher entropy quantiles, indicating that
even modest semantic exploration can exceed user tolerance under conservative
collaborative filtering.

Figure~\ref{fig:ncf_semantic} shows utility versus semantic
exploration for NCF. Utility increases sharply from very low exploration to moderate exploration, peaking around the lower–middle entropy quantiles.
After this point, utility fluctuates.
At very high exploration levels, utility does not improve further and even dips in some quantiles before partially recovering. This suggests
that increased model expressiveness delays but does not eliminate saturation.

Figure~\ref{fig:mostpop_movielens}  shows that
MostPopular is highly sensitive to exploration. Utility drops sharply at early
exploration levels, recovers briefly, and then declines again, highlighting that
popularity-based methods are particularly vulnerable to excessive novelty.

LightGCN appears flat in the aggregated comparison due to scale compression, but standalone visualization reveals meaningful within-model variation in utility across exploration levels.

Figure~\ref{fig:lightgcn} shows that Light-GCN does not exhibit a strong dip, it also fails to achieve meaningful
gains from increasing exploration. This response indicates saturation through insensitivity rather than degradation.
Taken together, these figures demonstrate that saturation does not require sharp
utility collapse.

\subsection{Marginal Effects of Exploration}

Aggregate utility curves can obscure when exploration ceases to provide
incremental benefit. To directly operationalize exploration saturation as defined
in Section~4, we analyze marginal changes in utility across ordered exploration
regimes. Specifically, we approximate the marginal effect of exploration as
$\Delta U(Q_k)$, measuring changes in average utility between successive
exploration quantiles.

\begin{figure}[t]
\centering
\includegraphics[width=\linewidth]{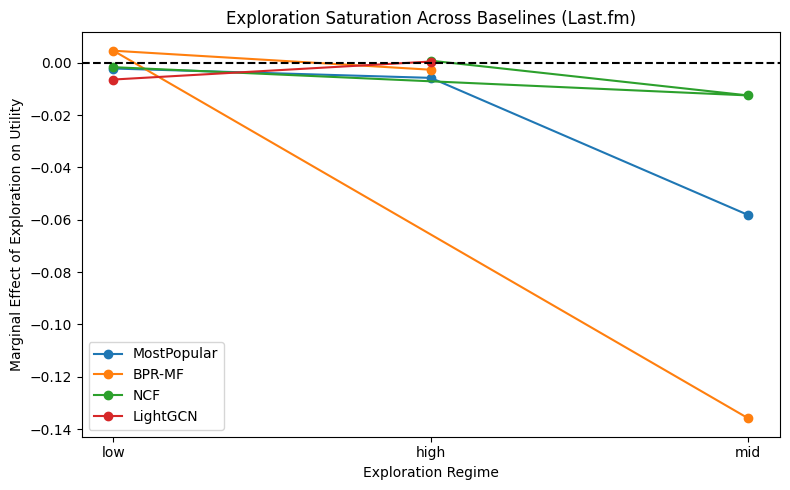}
\caption{Marginal effect of exploration on utility across recommendation models
on Last.fm.}
\label{fig:lastfm_marginal}
\end{figure}

\begin{figure}[t]
\centering
\includegraphics[width=\linewidth]{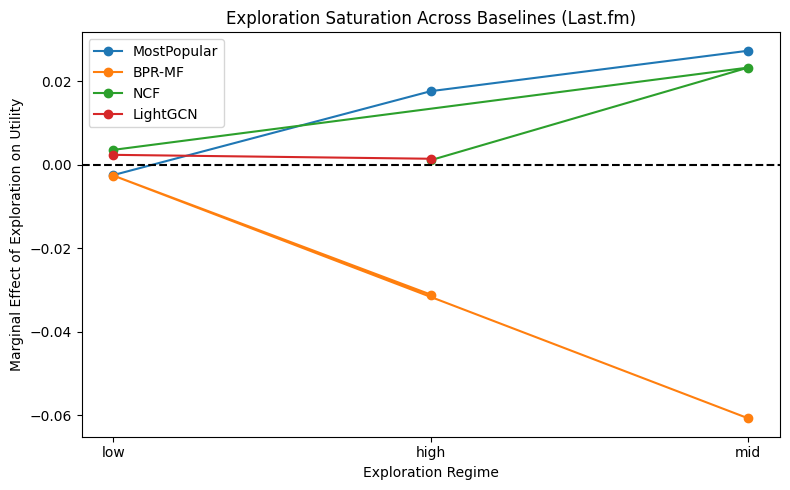}
\caption{Marginal effect of exploration on utility across recommendation models
on MovieLens-1M.}
\label{fig:movielens_marginal}
\end{figure}

Figure~\ref{fig:lastfm_marginal}
 shows marginal utility effects on Last.fm. At low
exploration levels, several models exhibit positive or near-zero marginal gains,
indicating that modest exploration can improve engagement. However, as
exploration increases, marginal utility becomes negative for BPR-MF and
MostPopular, satisfying the saturation condition and indicating that additional
exploration actively harms user utility. For NCF and LightGCN, marginal gains
diminish rapidly and converge toward zero, indicating saturation through
exhaustion of benefit rather than sharp decline.

Figure~\ref{fig:movielens_marginal}
shows a similar pattern on MovieLens-1M, though
with weaker magnitude. Marginal utility approaches zero across all models after
early exploration regimes, and remains non-positive at higher levels. Even in
the absence of strong negative effects, convergence toward zero marginal utility
indicates that exploration benefits are bounded and saturate quickly in this
denser domain.

Across both datasets, marginal analysis reveals that exploration should not be
treated as a monotonic objective. Saturation often emerges before observable
collapse in aggregate utility, manifesting instead as vanishing or unstable
marginal gains.

\subsection{User-Level Exploration Saturation}
Aggregate trends can mask substantial variation in how individual users respond
to exploration. We therefore analyze exploration saturation at the user level,
explicitly identifying user-specific saturation points based on marginal utility
changes.

Following our definition, we identify each user’s saturation point as the
earliest exploration regime $Q_k$ for which the marginal utility
$\Delta U_u(Q_k) \le 0$. This allows us to characterize not only whether
exploration saturates, but \emph{when} saturation occurs for different users.

\begin{figure}[t]
\centering
\includegraphics[width=\linewidth]{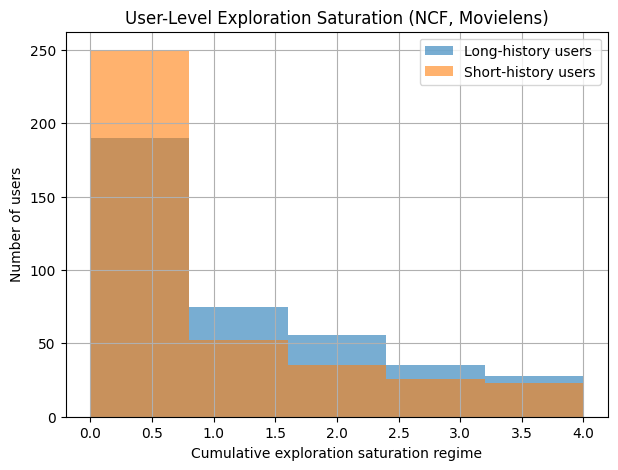}
\Description{Visualization of user-level exploration saturation regimes on MovieLens-1M.}
\caption{Distribution of user-level exploration saturation regimes on
MovieLens-1M, stratified by interaction history length.}
\label{fig:user_movielens}
\end{figure}

\begin{figure}[t]
\centering
\includegraphics[width=\linewidth]{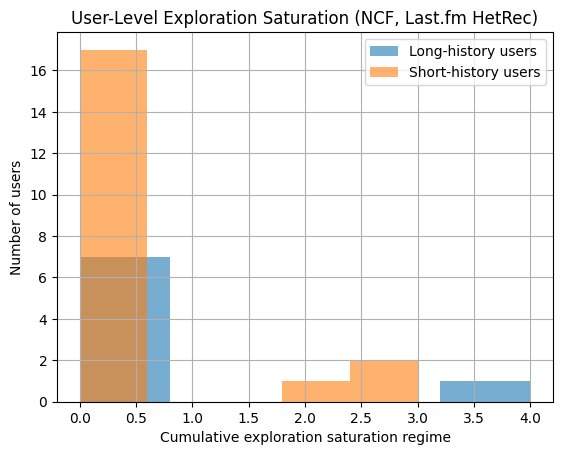}
\caption{User-level exploration saturation for NCF on Last.fm.}
\label{fig:user_lastfm}
\end{figure}
Figure~\ref{fig:user_movielens} shows the distribution of
saturation regimes for MovieLens-1M users, stratified by interaction history
length. Users with short interaction histories concentrate heavily in early
saturation regimes, indicating that they reach non-positive marginal utility
under relatively low levels of exploration. In contrast, users with longer
histories exhibit a broader distribution of saturation points, reflecting
greater tolerance for exploration.

Figure~\ref{fig:user_lastfm}
shows a similar pattern on Last.fm under
NCF. Due to higher sparsity and item turnover, many users reach saturation almost
immediately, while a smaller subset tolerates extended exploration. This
distribution highlights extreme heterogeneity in exploration tolerance within a
single recommendation model.

Taken together, these results demonstrate that exploration saturation is fundamentally user-dependent. Uniform exploration strategies, whether introduced implicitly by model behavior or explicitly through fairness and novelty interventions, impose unequal utility costs across users, over-penalizing users with low exploration tolerance while under-serving those capable of sustained exploration.

\subsection{Summary}

Across datasets, models, and analysis levels, exploration exhibits bounded
benefits. Utility gains saturate, fluctuate, or decline as exploration increases,
with substantial variation across users. These findings empirically demonstrate
that fairness and novelty-driven exploration must be treated as a constrained,
user-dependent process rather than a universally increasing objective.

\section{Discussion}

This work examines a common assumption in fairness-aware recommender systems that increasing exploration to improve fairness is always beneficial, or at least harmless, for users. Across multiple datasets, models, and analyses, our results show that this assumption does not hold in general. We find that exploration has bounded and non-monotonic effects on user utility, with clear saturation and diminishing returns, and that these effects vary substantially across users. In this section, we discuss the implications of these findings for fairness-aware recommendation, evaluation practices, and the design of user-centered recommender systems.

\subsection{Exploration Saturation Beyond Performance Collapse}

A central insight from our analysis is that exploration saturation does not
require a visible collapse in conventional performance metrics. In several
settings, particularly on MovieLens-1M and for more expressive models such as NCF
and LightGCN, utility curves flatten, oscillate, or exhibit increased variance
rather than a sharp decline. Crucially, these patterns should not be interpreted
as evidence that continued exploration is harmless or desirable.

From a behavioral perspective, flat or unstable utility indicates that
additional novelty fails to provide a reliable benefit to users. Even in the
absence of measurable performance degradation, excessive exploration can impose
cognitive burden, reduce perceived relevance, or introduce uncertainty in
recommendation quality. Such effects are difficult to capture with aggregate
offline metrics, yet they signal that exploration has exceeded a useful range.
Our findings, therefore, motivate a broader definition of exploration saturation
as the point at which marginal gains vanish or become unreliable, rather than
only when performance drops below a baseline.
This reinterpretation challenges evaluation practices that prioritize peak or
average utility while ignoring stability and marginal effects. In fairness-aware
contexts, such practices risk endorsing exploration regimes that appear
acceptable in aggregate but are already suboptimal or harmful for subsets of
users.

\subsection{Fairness as a User-Experienced Intervention}

A key conceptual contribution of this work is reframing fairness not solely as a
property of item exposure distributions, but as an intervention that is directly
experienced by users. Many fairness-aware recommendation approaches seek to
correct systemic biases by increasing exposure to under-represented or long-tail
items, often through mechanisms that promote novelty or diversity. While such
interventions can improve equity at the catalog or population level, they
necessarily alter the nature of content presented to individual users.
Our results show that this exploration pressure is not uniformly tolerated.
Users with shorter or less stable interaction histories consistently reach
exploration saturation earlier than users with richer histories. As a result,
uniform fairness or novelty promotion can impose disproportionate experiential
costs on certain users, even in the absence of explicit demographic attributes or
protected classes. This phenomenon represents a subtle but important form of
unequal user impact that is largely invisible to standard fairness metrics.
By exposing this heterogeneity, our findings suggest that fairness-aware systems
should consider not only who receives exposure, but also how that exposure is
experienced. 

\subsection{Limits of Static Fairness and Exploration Controls}

The observed variability in exploration saturation highlights fundamental
limitations of widely used fairness and exploration control mechanisms. Common
approaches, including fixed regularization strengths, global diversity
constraints, heuristic exposure caps, and offline hyperparameter tuning implicitly
assume that a single exploration setting can balance fairness and utility across
users, contexts, and stages of interaction.
Our results indicate that this assumption is overly simplistic. Even when such
controls are carefully tuned, they remain static and coarse-grained, unable to
respond to user-specific tolerance or temporal dynamics. Consequently, they may
overcorrect for bias in some cases while unnecessarily degrading experience in
others. This critique is not meant to dismiss existing methods, but rather to
clarify the limits of what static controls can achieve when applied to a
fundamentally user-dependent phenomenon.
Our contribution is diagnostic rather than prescriptive. We do not
advocate abandoning existing fairness techniques; instead, we show that their
effects should be interpreted through the lens of user experience and
exploration saturation, rather than solely through aggregate fairness or accuracy metrics.
\subsection{Implications for Adaptive and Human-Centered Recommendation}
Although adaptive recommendation strategies are widely studied, most existing
approaches adjust exploration strength indirectly through parameter updates,
reward optimization, or predefined control rules. Such methods typically assume
that exploration remains beneficial until objective performance metrics begin
to degrade, and rarely model user tolerance for exploration explicitly.
Our findings suggest that this assumption is incomplete. Exploration saturation
often emerges before observable performance collapse, manifesting instead as
utility plateaus, increased instability, or unequal user impact. As a result,
adaptive systems that rely exclusively on objective feedback may respond too
late, or fail to recognize when exploration has ceased to provide value for
specific users.
Rather than proposing a new adaptive algorithm, our work provides empirical
evidence that adaptivity must be informed by an understanding of user-specific
exploration limits. Exploration saturation serves as a diagnostic lens for
identifying when fairness-driven novelty is beneficial, when it is neutral, and
when it becomes counterproductive. In this sense, our analysis complements
existing adaptive methods by clarifying \emph{what} should be adapted,
\emph{for whom}, and \emph{why}.

From a human-centered perspective, these results underscore the importance of
treating fairness interventions as user-experienced processes rather than purely
distributional controls. By grounding fairness decisions in observed user
responses, future systems can better balance equity objectives with sustainable
user engagement, without assuming that increased novelty is universally
desirable.
\subsection{Design Implications for Fairness-Aware Recommendation}
First, fairness-driven exploration should not be treated as a uniform intervention.
Our results demonstrate that users differ substantially in their tolerance for
exploration, with users possessing shorter or less stable interaction histories
reaching saturation significantly earlier than others. As a result, global fairness
or novelty controls—such as fixed regularization strengths, static diversity
constraints, or uniform re-ranking strategies—risk over-penalizing certain users
while providing limited additional benefit to others.
Second, exploration pressure should be monitored using marginal and stability-based
signals rather than aggregate utility alone. Exploration saturation often emerges
before observable performance collapse, manifesting instead as diminishing marginal
returns or increased variability in utility. This suggests that adaptive systems
should consider early warning signals—such as vanishing marginal gains or unstable
user responses—as indicators that additional fairness-driven exploration may no longer
be beneficial for a given user.
Finally, our findings motivate user-aware exploration control mechanisms that adapt
the amount of exploration applied to each user, rather than solely adjusting item
relevance. Such mechanisms could condition exploration strength on behavioral signals
including interaction history length, consistency of engagement, or observed
sensitivity to novelty. Importantly, these adaptations need not introduce new fairness
objectives, but rather reinterpret existing fairness interventions through the lens
of user experience and tolerance.
Together, these implications suggest that fairness-aware recommendation should move
beyond static, population-level controls toward approaches that treat exploration as
a personalized, bounded, and user-experienced process.

\section{Conclusion}
In this work, we examined a fundamental yet underexplored assumption in
fairness-aware recommender systems: that increasing exploration in the service
of fairness is uniformly beneficial to users. Through a systematic empirical
study across datasets, models, and user groups, we showed that this assumption
does not generally hold. Instead, exploration exhibits bounded and
user-dependent benefits, with clear evidence of saturation, diminishing returns,
and heterogeneous tolerance across users.
Our findings highlight that fairness-driven novelty and diversity interventions
should be understood not only as distributional corrections, but as user-experienced processes that can impose varying cognitive and relevance
costs. Importantly, exploration saturation often emerges before observable
performance collapse, manifesting as utility plateaus or instability rather than
sharp declines. This suggests that conventional evaluation practices may
overlook critical signals indicating when exploration has ceased to provide
value.

\section{Limitations and Future Work}
This study focuses on empirically characterizing exploration saturation through measurement and analysis. Using controlled offline experiments on two widely used movie and music datasets with distinct interaction patterns, we evaluate exploration behavior across models and settings. Our findings provide guidance for the design and evaluation of adaptive and fairness-aware recommender systems. Future work includes extending the analysis to additional domains, conducting online and user-centered evaluations, and developing adaptive strategies that dynamically adjust exploration based on user behavior.

\section{Ethical \& Human Subjects Considerations}
This study relies exclusively on publicly available datasets (MovieLens and Last.fm) that contain anonymized user--item interaction logs. No new data were collected, and no direct interaction with human participants was conducted. As a result, ethical review or informed consent was not required for this work. Both datasets were originally released for research purposes and do not contain personally identifiable information.

\bibliographystyle{ACM-Reference-Format}
\bibliography{references}

\end{document}